\documentstyle[preprint,aps]{revtex}
\input psfig.tex
\begin{document}

\title{Domains of Disoriented Chiral Condensate} 

\author{R.~D.~Amado and Yang Lu} 
\address{Department of Physics,
University of Pennsylvania, 
Philadelphia, PA 19104}

\date{\today}

\maketitle

\begin{abstract}
The probability distribution of neutral pion fraction from 
independent domains of disoriented chiral condensate is 
characterized. 
The signal for the condensate is clear for a small number 
of domains but is greatly reduced for more than three.
\end{abstract}


There has been much interest recently in the possible
production of a disoriented chiral condensate in high
energy hadronic or heavy ion collisions \cite{lots,A&R,A&K}.
It has been proposed that this  condensate can be modeled as a
region of coherent classical pion field described either
in the linear or non-linear sigma model. These studies 
predict a striking probability distribution for the neutral
pion fraction, $f$, coming from the condensate,
\begin{equation}
f=\frac{n_0}{n_{total}},
\end{equation}
where $n_0$ is the number of observed neutral pions. 
If the field is constant in the formation region,
one finds a probability distribution of
\begin{equation}
 P(f) = \frac{1}{2 \sqrt{f}}.
\end{equation}
This is markedly different from the standard  statistical distribution 
which, for large $n_{total}$, we expect to be
\begin{equation}
P(f) = \delta (f-\frac{1}{3}).
\end{equation}
Recently Anselm and Ryskin have shown \cite{A&R} that
a region of classical pion field varying rapidly in 
space and time will also lead to an easily recognized
and characteristic 
probability distribution for the neutral fraction. 
Although in this case that distribution is not easily
given in closed form and is not quite so dramatic as 
for the uniform field case.

Previous studies have concentrated on a single classical
field region or single domain of coherent condensate. 
It is likely that, particularly in a heavy ion
collision, more that one domain or
 region of chiral condensate
will form with an independent ``direction" for the chiral
condensate in each domain.  In that 
case the observed neutral pion fraction, $f$,
will be an average over regions. In this note we discuss the 
probability of neutral pion fraction that emerges from averaging
over independent production domains both for the case of regions
of constant field and for regions of rapidly varying field as 
given by Anselm and Ryskin.  As we average over many regions the
probability distribution will tend to the statistical one of Eq.~(3),
but the physically interesting case is more likely one of a only
a few domains. 

We wish to average $f$ over N independent regions. In each single
region we take the probability of $f$ to be given by $P_1(f)$ and 
for simplicity we take the N regions to have equal weight. The 
probability of finding neutral fraction $f$ averaged over the N 
regions is given by
\begin{equation}
P_N(f) = \int df_1 ....df_N \delta\left(f-\frac{f_1+f_2+...f_N}{N}
\right)
    P_1(f_1) P_1(f_2).....P_1(f_N).
\end{equation}
This can be transformed into a recursion relation,
\begin{equation}
P_N(f) = \frac{N}{N-1} \int P_{N-1} \left(\frac{N f -f_N}{N-1}\right) 
P_1(f_N) d f_N.
\end{equation}
This relation is particularly helpful in computing $P_N$ stepwise in N.
Note that the recursion relation guarantees that the average value
of $f$ is the same for any N.  The average is 1/3 both for the constant
field case and for the rapidly varying field case studied by Anselm and
Ryskin.  

\begin{figure}
\centerline{\hbox{
\psfig{figure=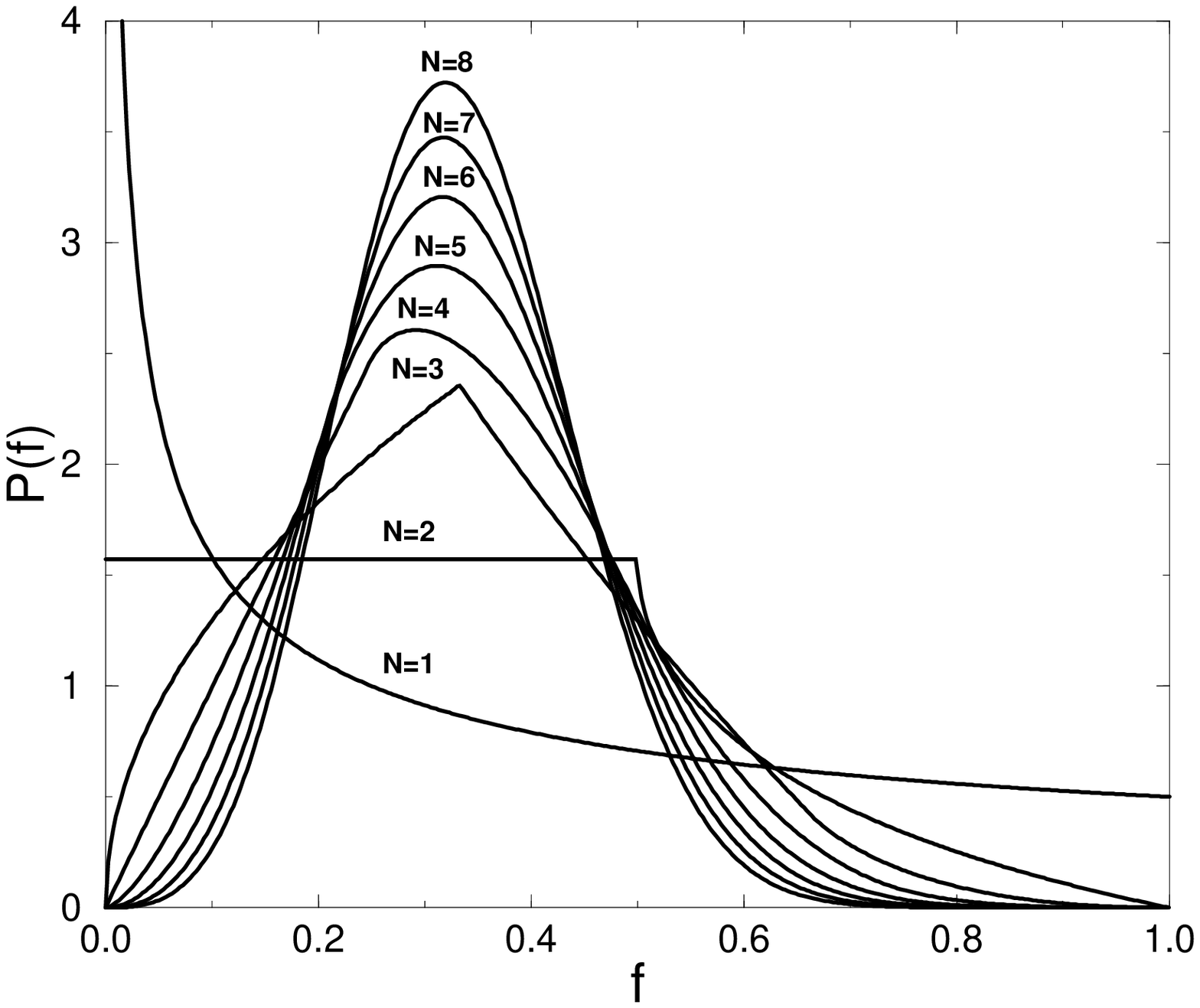,height=3.7in}
}}

\caption{
Probability density, $P(f)$, for neutral pion fraction $f$, with
multiple random domains of static disoriented chiral condensates.
Number of domains ranges from 1 to 8.
}\label{Fig.1}
\end{figure}

First let us consider the case of uniform pion field in each domain.
In this case $P_1(f) $ is given by Eq.~(2).  For two domains the probability
can be found analytically and we find
\begin{equation}
P_2(f) = \frac{\pi}{2}
\end{equation} 
for $f<\frac{1}{2}$ and
\begin{equation}
P_2(f) = \frac{\pi}{2} -2 \arccos (\frac{1}{ \sqrt{2f}})
\end{equation} 
for $f>\frac{1}{2}$. This result was previously calculated in \cite{A&K}.
Beyond two regions the integrals are most easily done numerically. 
The results of $N=1....8$ are shown in Fig.~\ref{Fig.1}.  
The approach to
a gaussian distribution centered at $1/3$ is clear.  
This is a consequence of the central limit theorem. The standard deviation 
of the distribution is proportional 
to $1/\sqrt{N}$ and decreases as $N$ gets larger, 
finally reducing the distribution at very large $N$
to a delta function at $1/3$.  What is also clear is that it
would require a high statistics experiment to distinguish four
or more domains from the gaussian of incoherent pion production.

\begin{figure}
\centerline{\hbox{
\psfig{figure=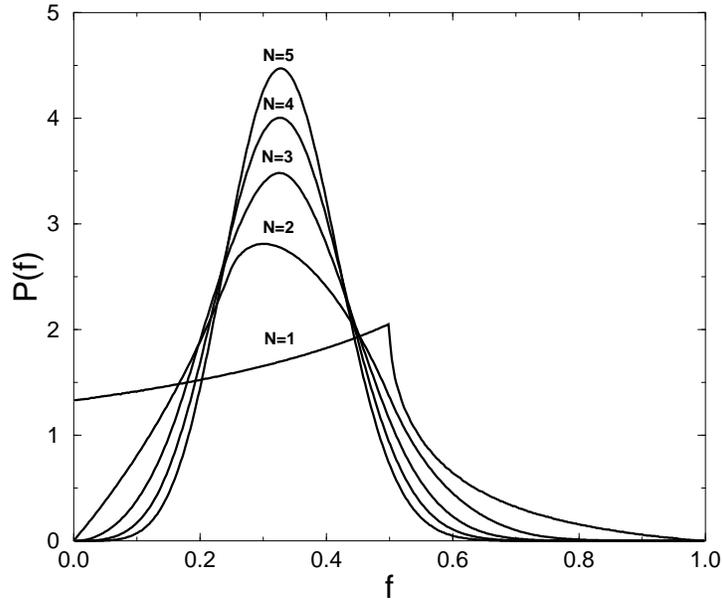,height=3.7in}
}}

\caption{
Probability density, $P(f)$, for neutral pion fraction $f$, with
multiple random domains of rapidly time-dependent 
disoriented chiral condensates.
Number of domains ranges from 1 to 5.
}\label{Fig.2}
\end{figure}

The case of rapid variation of the pion field in each domain is 
shown in Fig.~\ref{Fig.2}.  For $N=1$ we have the result of Anselm and 
Ryskin calculated numerically.  
This result assumes the variation of the field can be averaged over.
We then use the recursion relation to obtain
the higher $P_N$ up to $N=5$.  The approach to a gaussian distribution
is again clear.  The experimental challenge with many domains is
more daunting in this case, since even two or three domains will be
very difficult to distinguish from the random case.

One can mix domains  from the Anselm Ryskin case with constant domains
or complicate things more by giving unequal weight to domains.  It is
clear that so long as there are fewer than three domains it should
be possible to extract a signal of the chiral condensate, but if there
are more than three it will be difficult.  

We have shown how the neutral fraction of pions from independent
domains of chiral condensate can be characterized.  It should be 
possible to see a signal of the condensate so long as the number of
domains is not large, even if that number varies from collision to 
collision, but more than one domain will make the signal less dramatic.

This work was supported in part by a grant from the National Science Foundation.


\end{document}